\documentclass[aps,prc,twocolumn,floatfix,nofootinbib,superscriptaddress]{revtex4-1}
\usepackage{float}
\usepackage[T1]{fontenc}
\usepackage[utf8]{inputenc} 
\usepackage{amsmath,amsfonts,amssymb}
\usepackage{graphicx}
\usepackage{epstopdf} 
\usepackage{xcolor}
\usepackage{hyperref} 
\hypersetup{colorlinks=true,linkcolor=blue,citecolor=blue,urlcolor=blue}
\usepackage{bm}

\begin{document}

\title{Bayesian Smooth-Fit Extrapolation of the $^{12}\mathrm{C}+{}^{12}\mathrm{C}$ Astrophysical $S$ Factor}

\author{A.~M.~Mukhamedzhanov}
\affiliation{Texas A\&M University, College Station, Texas 77843, USA}

\date{\today}

\begin{abstract}
A Bayesian analysis of the astrophysical \(S\) factor for the
\(^{12}{\rm C}+{}^{12}{\rm C}\) fusion reaction is presented using
available low-energy information at carbon--carbon relative energies
\(E <3.5~{\rm MeV}\), including direct measurements and recent
inverse-kinematics data. The goal of the global Bayesian fit is not to
reproduce the local resonance-by-resonance structure of the
\(^{12}{\rm C}+{}^{12}{\rm C}\) system, but rather to constrain the
smooth global component of \(S^{*}(E)\) using the direct and
inverse-kinematics constraints.
A conservative uncertainty model is adopted in order to combine
heterogeneous datasets with different systematic uncertainties and
normalizations. The inference is performed for
\(y(E)=\log_{10}S^{*}(E)\), which reduces the dynamic range of the data
and leads to approximately Gaussian uncertainties. The function
\(y(E)\) is represented by a quadratic polynomial, and the posterior
distribution of its coefficients is determined. Propagation of this
posterior gives the median smooth curve and the corresponding \(68\%\)
posterior interval for \(S^{*}(E)\).

The resulting posterior \(S^{*}(E)\) lies systematically below the
traditional FCZ75 reference normalization over the energy interval
considered.  For example, at \(E=1.5~{\rm MeV}\) the median posterior
value is
\(S^{*}(1.5~{\rm MeV})=1.20\times10^{16}~{\rm keV\,b}\), whereas the
Fowler--Caughlan--Zimmerman reference value is
\(S^{*}_{\rm FCZ75}=3.0\times10^{16}~{\rm keV\,b}\)
[W.~A. Fowler, G.~R. Caughlan, and B.~A. Zimmerman, Annu. Rev. Astron.
Astrophys. \textbf{13}, 69 (1975)].  Thus, the updated posterior favors
a lower smooth low-energy trend and does not support a sharp increase of
\(S^{*}(E)\) as the energy decreases.

The astrophysical consequence of the analysis is assessed through the
thermonuclear reaction rate \(N_A\langle\sigma v\rangle\).  The resulting
median rate is lower than the CF88 analytic rate
[G.~R. Caughlan and W.~A. Fowler, At. Data Nucl. Data Tables
\textbf{40}, 283 (1988)] over the temperature interval considered, with
\(R_{\rm present}/R_{\rm CF88}\simeq0.33\)--\(0.46\) for
\(0.3\leq T_9\leq1.2\).  At higher temperatures, where the Gamow window
extends above the upper end of the adopted integration interval, the
calculated rate should be regarded as a truncated rate rather than a
complete stellar rate.
\end{abstract}

\maketitle

\section{Introduction}
\label{secintro}

The \(^{12}{\rm C}+{}^{12}{\rm C}\) fusion reaction plays a central role
in the late stages of stellar evolution and in explosive astrophysical
scenarios.  Its low-energy astrophysical factor governs carbon burning
in massive stars and contributes to the modeling of Type~Ia supernovae
and superbursts.  In particular, one proposed channel for Type~Ia
supernovae is the merger of two white dwarfs~\cite{Mori}, for which the
outcome depends sensitively on the \(^{12}{\rm C}+{}^{12}{\rm C}\)
fusion rate.  Reliable extrapolations of the astrophysical \(S\) factor
to carbon--carbon energies \(E\lesssim 2~{\rm MeV}\) are therefore
required to constrain reaction rates under stellar conditions; see, for
example, the recent comprehensive review~\cite{Chieffi}.  Direct
measurements in this regime are extremely challenging because of the
rapidly decreasing cross section and increasing background.  As a
result, most direct experimental information is concentrated at higher
energies and must be supplemented by indirect or inverse-kinematics
constraints to anchor the sub-barrier behavior.

Over the past decades, a diverse experimental landscape has emerged for
the \(^{12}{\rm C}+{}^{12}{\rm C}\) fusion reaction.  Early direct
measurements were reported in
Refs.~\cite{Mazarakis,Becker,Spillane,Aguillera}, followed by more
recent direct studies by the Notre Dame group and collaborators
\cite{Tan20,Tan24} and by the STELLA collaboration
\cite{Fruet,Nippert}.  These direct measurements are largely restricted
to energies \(E_{\rm cm}\gtrsim 2.1~{\rm MeV}\), i.e., above the main
Gamow-window region relevant for hydrostatic carbon burning.

In addition to traditional direct measurements, two major experimental
developments have significantly advanced the understanding of the
low-energy behavior of the \(^{12}{\rm C}+{}^{12}{\rm C}\) fusion
reaction.  The Trojan Horse Method (THM) analysis~\cite{Tumino} provided
an indirect extension into the sub-barrier region, while the more recent
inverse-kinematics measurements of Ref.~\cite{NanWang} provided direct
low-energy constraints through detailed balance.  The inverse-kinematics
data of Ref.~\cite{NanWang} establish an experimentally constrained band
for the fusion \(S\) factor in the energy region most relevant for
astrophysical extrapolation.  The initial plane-wave THM analysis
suggested a pronounced low-energy enhancement of the astrophysical
\(S\) factor.  Subsequent inclusion of Coulomb--nuclear interactions in
the THM analysis~\cite{MukTHM} significantly modified this behavior,
suppressing the enhancement and bringing the extracted \(S\) factor into
qualitative agreement with the inverse-kinematics results of
Ref.~\cite{NanWang}.

The most advanced microscopic calculations of the low-energy
\(S\) factor for the \(^{12}{\rm C}+{}^{12}{\rm C}\) fusion reaction
have been reported in Refs.~\cite{Kimura21,Kimura23}.  These
calculations do not support either the hindrance pattern proposed in
Ref.~\cite{Jiang} or the sharp low-energy enhancement inferred from the
original THM analysis~\cite{Tumino}.  A recent re-examination of the
\(^{12}{\rm C}+{}^{12}{\rm C}\) fusion reaction~\cite{Hagino} also found
no compelling evidence for a fusion-hindrance pattern at low energies,
further underscoring the uncertainties associated with extrapolations
based on specific hindrance assumptions.  However, a recent multichannel
Resonating Group Method analysis in Ref.~\cite{Descouvemont26} suggests
that the low-energy \(S^{*}\) factor\footnote{For heavy-ion fusion
reactions it is customary to introduce the modified astrophysical factor
\[
S^{*}(E)=S(E)\exp(0.46E),
\]
where \(E\) is in MeV.  Equivalently, the fusion cross section may be
written as
\[
\sigma(E)
=
\frac{S^{*}(E)}{E}
\exp\left[-2\pi\eta(E)-0.46E\right],
\]
where \(\eta(E)\) is the Sommerfeld parameter.  The additional
\(\exp(0.46E)\) factor is a conventional heavy-ion modification that
reduces the rapid energy dependence of the ordinary astrophysical
factor and facilitates comparison of different
\(^{12}{\rm C}+{}^{12}{\rm C}\) datasets.} may exhibit hindrance-type
behavior.  Thus, there is significant disagreement in theoretical
descriptions of the low-energy astrophysical factor.

The central challenge is therefore to combine heterogeneous datasets in
a statistically consistent manner, accounting for their different energy
coverages, normalization uncertainties, and systematic assumptions, while
quantifying the resulting uncertainty in a form suitable for low-energy
extrapolation and thermonuclear reaction-rate calculations.

Two remarks clarify the scope of the present analysis.  First, the
Bayesian global fit is designed to extract a smooth global trend of
\(S^{*}(E)\), rather than to provide a resonance-by-resonance
parametrization of the \(^{12}{\rm C}+{}^{12}{\rm C}\) system.  The
adopted data contain pronounced local resonance structures, and these
structures enter the likelihood through the measured values of
\(S^{*}(E)\).  They therefore influence the fitted normalization, slope,
and curvature of the smooth posterior.  However, the present fit does
not assign separate resonance energies, widths, or interference phases
to individual structures.  The posterior curve should therefore be
interpreted as a smooth average constrained by resonance-affected data,
not as a local model of individual resonances.

Second, the thermonuclear reaction rate is constructed from a hybrid
input.  In the low-energy region, the direct inverse-kinematics data of
Ref.~\cite{NanWang} are used so that the experimentally observed
resonance structures are retained explicitly in the Maxwellian rate
integral.  At higher energies, this input is matched to the Bayesian
smooth posterior.  Thus, the low-energy resonances observed in
Ref.~\cite{NanWang} enter the reaction rate directly, while the
higher-energy contribution is supplied by the statistically constrained
smooth global component.

The Coulomb--nuclear-renormalized THM data are shown for comparison but
are not included as an independent constraint in the adopted Bayesian
likelihood.  This avoids double counting of overlapping low-energy
resonance structures already represented by the inverse-kinematics data
of Ref.~\cite{NanWang} and reflects the remaining model dependence of
the renormalized THM extraction.  Presenting a smooth global trend is
standard in \(^{12}{\rm C}+{}^{12}{\rm C}\) fusion studies and is widely
used in reaction-rate evaluations, including classic compilations such
as Caughlan--Fowler~\cite{Caughlan} and more recent smooth
extrapolation approaches, including hindrance-motivated
parameterizations~\cite{Jiang,DiazTorres,Esbensen,Beck}.

To this end, a Bayesian smooth-fit framework is adopted in which
\[
y(E)=\log_{10}S^{*}(E)
\]
is represented by a low-order polynomial in the carbon--carbon relative
energy \(E\).  Working in logarithmic space reduces the dynamic range
spanned by the experimental data and allows heterogeneous datasets to be
combined in a statistically stable manner.  A conservative uncertainty
prescription is adopted to account for the normalization and systematic
differences among the datasets.

Within this framework, the selected datasets contribute coherently to the
inference of the model parameters, and their uncertainties are propagated
through the covariance matrix.  The Bayesian treatment yields a posterior
probability distribution for the fit parameters and for derived
quantities.  This enables the construction of posterior low/median/high
curves for the smooth astrophysical factor.  The median curve represents
the posterior median of the smooth component, while the lower and upper
curves define the adopted posterior interval.

The quantity \(S^{*}(1.5~{\rm MeV})\) is used only as a compact
diagnostic of the smooth extrapolated component.  It should not be
interpreted as a local resonance-resolved observable, since the
\(^{12}{\rm C}+{}^{12}{\rm C}\) fusion cross section contains pronounced
resonant structures at low energy.  The primary physical output of the
analysis is instead the thermonuclear reaction rate
\(N_A\langle\sigma v\rangle\), obtained by propagating the adopted
posterior through the Maxwellian rate integral.

The resulting posterior curves and reaction-rate band provide a compact,
transparent, and reproducible summary of the global information currently
available from direct measurements and the inverse-kinematics constraint of
Ref.~\cite{NanWang}.  The upper bound of \(3.5~{\rm MeV}\) is chosen to include
the highest energies at which multiple datasets overlap while remaining within
the regime where a smooth parametrization of
\(\log_{10}S^{*}(E)\) is appropriate.  Reaction rates obtained from this
finite integration interval should therefore be interpreted as complete only
for temperatures for which the dominant Gamow window lies within the adopted
energy range.

\section{Data Sets and Uncertainty Assignments}
\label{secdata}

The analysis uses direct measurements together with the
inverse-kinematics data of Ref.~\cite{NanWang} to constrain the smooth
component of the modified astrophysical factor \(S^{*}(E)\).  The
Coulomb--nuclear-renormalized THM data are not included as an independent
constraint in the adopted Bayesian likelihood, but are retained as an
external comparison.  The individual datasets and uncertainty
assignments are described below.

\subsection{Experimental input}

The direct data include recent measurements performed by the Notre Dame
group and collaborators~\cite{Tan20,Tan24}, covering
\(E_{\rm cm}\gtrsim 2.6~{\rm MeV}\), recent measurements by the STELLA
collaboration~\cite{Fruet,Nippert}, covering
\(E_{\rm cm}\gtrsim 2.3~{\rm MeV}\), and earlier direct measurements
reported in Refs.~\cite{Mazarakis,Becker,Spillane,Aguillera}.

Through detailed balance, the inverse-kinematics measurements of
Ref.~\cite{NanWang} provide a direct determination of the
\(^{12}{\rm C}+{}^{12}{\rm C}\) fusion \(S^{*}\) factor at low energies
and define an experimentally constrained band in the region where resonant
structures are important.  These data are included in the adopted
Bayesian analysis together with the direct measurements.

The Coulomb--nuclear-renormalized THM data of Ref.~\cite{MukTHM} are
shown in Fig.~\ref{figpost_linS} as an external comparison, but are not
included as an independent constraint in the adopted Bayesian likelihood.
As discussed in Sec.~\ref{secintro}, this avoids double counting of
low-energy resonance information already constrained by the
inverse-kinematics data of Ref.~\cite{NanWang} and reflects the residual
model dependence of the renormalized THM extraction.

\subsection{Uncertainty assignments}

To maintain a transparent and reproducible statistical scheme, an
effective uncertainty is assigned to each experimental point.  The
assigned uncertainty retains the quoted point-to-point statistical
component and includes an additional conservative fractional component
to account for dataset-to-dataset normalization differences and other
residual systematic effects.  This prescription is applied uniformly to
the heterogeneous datasets entering the adopted global smooth fit, with
the exception of the inverse-kinematics data of Ref.~\cite{NanWang}, for
which the published energy-dependent upper and lower band is used
directly.

I emphasize at the outset that the Bayesian inference is performed for
the logarithmic variable
\[
y(E)=\log_{10}S^{*}(E),
\]
rather than for \(S^{*}(E)\) itself.  Therefore, the effective
uncertainties entering the likelihood are uncertainties in
\(\log_{10}S^{*}(E)\).  This choice is advantageous because it reduces
the large dynamic range of the astrophysical factor and leads to a more
nearly Gaussian description of the experimental scatter.

\subsection{Adopted uncertainty assignment}

For each experimental point, the quoted point-to-point uncertainty is
retained as the statistical component.  In addition, an overall
fractional component is included to represent residual systematic
effects, such as normalization shifts, efficiency corrections,
background subtraction uncertainties, and other experiment-dependent
effects that are not fully captured by counting statistics alone.
Accordingly, for all datasets except Ref.~\cite{NanWang}, the effective
relative uncertainty assigned to a point at energy \(E_i\) is taken as
\begin{equation}
\delta_i^{\rm eff}
=
\sqrt{\delta_{i,\rm stat}^{\,2}+\delta_{\rm sys}^{\,2}},
\qquad
\delta_{\rm sys}=0.25 .
\label{eq:delta_eff}
\end{equation}
The corresponding uncertainty in \(\log_{10}S^{*}(E)\) is then obtained
by standard error propagation.  If \(\delta_i^{\rm eff}\) denotes the
effective fractional uncertainty in \(S^{*}(E)\), the uncertainty
assigned in log space, see Appendix~\ref{sec:logsigma_to_fraction}, is
\begin{equation}
\sigma_{y,i}= \log_{10}(1+ \delta_i^{\rm eff})
\approx
\frac{\delta_i^{\rm eff}}{\ln 10}.
\label{eq:sigma_y_from_delta}
\end{equation}

The rationale for introducing the additional fractional term
\(\delta_{\rm sys}=0.25\) is the following.  Many datasets for
\(^{12}{\rm C}+{}^{12}{\rm C}\) report comparatively small point-to-point
statistical uncertainties, whereas the dominant differences among
experiments at similar energies often arise from dataset-to-dataset
normalization differences and other residual systematic effects.  If the
global smooth regression were weighted using only the quoted statistical
errors, the fit would be dominated by subsets of points with unusually
small \(\delta_{i,\rm stat}\), even when other independent measurements
in the same energy interval are visibly displaced from them.  In that
case the inferred posterior band would become unrealistically narrow and
would effectively collapse toward a thin line, giving the misleading
impression that the smooth global trend is known much more precisely
than is actually justified by the experiment-to-experiment spread.
Equation~(\ref{eq:delta_eff}) is therefore adopted as a simple and
uniform way of retaining the quoted point-to-point uncertainties while
also accounting for residual systematic effects that are otherwise left
unmodeled.

For the inverse-kinematics data of Ref.~\cite{NanWang}, the published
results are given not as individual statistical error bars of the same
type as in the other datasets, but rather as an adopted central curve
together with an energy-dependent upper and lower band.  In this case,
the relative uncertainty is assigned directly from that band according
to
\begin{equation}
\delta_i =\max\!\left[
 \frac{S_i^{\rm hi}(E_i)-S_i^{\rm lo}(E_i)}
     {2\,S_i^{\rm ad}(E_i)},
\;0.05
\right],
\label{eq:delta_nanwang}
\end{equation}
where \(S_i^{\rm hi}\), \(S_i^{\rm lo}\), and \(S_i^{\rm ad}\) denote
the upper, lower, and adopted central values, respectively.  A minimum
\(5\%\) uncertainty is imposed in order to avoid an artificially narrow
constraint from this dataset in regions where the published band becomes
very small.

\subsection{Interpretation}

The combined adopted dataset is used to constrain a smooth representation of
\(\log_{10}S^{*}(E)\) within the Bayesian framework described above.  The
adopted uncertainty assignments enter directly into the likelihood and
therefore play a central role in determining the width of the global posterior
band.  Accordingly, the quoted credible intervals should be interpreted not
only as a reflection of the available experimental information, but also as a
reflection of the conservative assumptions used to describe the spread of the
data and the remaining systematic differences between the datasets.

\subsection{Quadratic Representation}
\label{QuadraticRep}

As discussed in Sec.~\ref{secintro}, working with
\(\log_{10} S^{*}(E)\) substantially reduces the dynamic range spanned by
the experimental data.  It also gives a natural statistical representation of
the uncertainties, because many normalization and systematic effects enter
approximately as relative, or multiplicative, uncertainties in \(S^{*}(E)\).
Taking the logarithm converts such multiplicative variations into additive
deviations in \(y(E)=\log_{10}S^{*}(E)\).  This representation facilitates a
statistically stable combination of datasets obtained using different
experimental techniques.

The logarithm of the modified astrophysical factor is assumed to be adequately
represented over the energy range
\[
0.5 \le E \le 3.5~\mathrm{MeV}
\]
by a quadratic polynomial,
\begin{align}
y(E;\mathbf{a})
&=\log_{10} S^{*}(E;\mathbf{a})
\nonumber\\
&= a_0 + a_1 (E-E_0) + a_2 (E-E_0)^2 ,
\label{eqpoly1}
\end{align}
where \(E_0\) is a fixed pivot energy chosen within the fitted range.
In the present analysis, \(E_0=1.5~\mathrm{MeV}\) is adopted.  Centering
the polynomial at \(E_0\) improves numerical stability and reduces
correlations among the fitted coefficients.  The resulting posterior
distributions are found to be insensitive to moderate changes of the pivot
energy within the fitted interval.

For each data point \(i\) with measured or derived central value \(S_i^{*}\)
and fractional uncertainty \(\delta_i\), the corresponding uncertainty in
\(\log_{10} S^{*}\) is taken as
\begin{align}
\sigma_i = \frac{\delta_i}{\ln 10},
\label{eqsigma_log}
\end{align}
which follows from standard error propagation.

A weighted least-squares fit in \(\log_{10} S^{*}(E)\), with weights
\(w_i=1/\sigma_i^2\), is performed to determine the coefficient vector
\(\mathbf{a}=(a_0,a_1,a_2)\) together with its covariance matrix
\(\mathbf{C}\).

Because the model is linear in the coefficients \(a_0\), \(a_1\), and
\(a_2\), and the experimental uncertainties are treated as Gaussian in
\(\log_{10} S^{*}(E)\), the statistical uncertainties of the fitted
coefficients are themselves Gaussian.  Under these conditions, the
covariance matrix \(\mathbf{C}\) fully specifies the width and shape of the
resulting parameter distribution and provides a Gaussian approximation to the
posterior distribution of the model parameters \(\mathbf{a}\).

Parameterizations equivalent to a low-order polynomial in \(\ln S(E)\),
that is, \(S(E)\) expressed as an exponential with linear or quadratic terms
in the exponent, have been used previously in the literature; see, e.g.,
Refs.~\cite{Yakovlev2010,Afanasjev2012}.  The present work differs in that
the smooth log-space representation is embedded into a Bayesian framework and
the global parameter uncertainties are propagated to \(S^{*}(E)\) through
Monte Carlo sampling of the full coefficient covariance, producing a median
curve and a central \(68\%\) credible band.

No prior theoretical constraints are imposed on the coefficients
\(a_0\), \(a_1\), and \(a_2\) beyond the requirement of a smooth functional
form.  In particular, neither the overall normalization nor the slope or
curvature of \(\log_{10} S^{*}(E)\) is restricted by hand.  The fitted
parameter values and their uncertainties are therefore determined by the
adopted experimental data and their assigned errors.

\subsection{\texorpdfstring{Justification for the Quadratic Regression Model}
{Justification for the Quadratic Regression Model}}

A common concern is the use of a quadratic polynomial to represent
\(\log_{10} S^{*}(E)\) in the presence of resonant structures in the data.
The following points address this concern:
\begin{enumerate}
    \item The quadratic model is applied over a restricted energy interval,
    \(0.5\le E\le 3.5~\mathrm{MeV}\), within which it is used only to describe
    the smooth global component of \(\log_{10}S^{*}(E)\).

    \item The polynomial parametrization is not intended as a physical model
    of the reaction mechanism; it serves as a regression model for the smooth
    background trend.

    \item Resonance-affected data points are not discarded.  Their effect is
    retained through the likelihood and contributes to the fitted global
    normalization, slope, and curvature according to their assigned
    uncertainties.  The resulting curve should therefore be interpreted as a
    smooth trend averaged over the resonance-affected data, not as a
    resonance-by-resonance description.

    \item The inclusion of higher-order polynomial terms does not improve the
    Bayesian evidence and leads to additional coefficients that are
    statistically insignificant.

    \item Robustness checks confirm that the extracted value of
    \(S^{*}(1.5~\mathrm{MeV})\) is stable under cubic or quartic extensions of
    the polynomial, with deviations well below the quoted credible intervals.
\end{enumerate}
\section{Bayesian formalism}
\label{BayesForm}

\subsection{Bayes' theorem}

The statistical inference problem addressed in this work is formulated
within a Bayesian framework. Given a set of experimental data \(D\),
Bayes' theorem relates the posterior probability density to the
likelihood and prior according to
\begin{align}
P(\theta\mid D)=
\frac{\mathcal L(D\mid\theta)\,P(\theta)}{P(D)},
\label{eqbayes_theta}
\end{align}
where \(P(\theta\mid D)\) is the posterior probability density for the
parameter \(\theta\), \(P(\theta)\) is the prior distribution, and
\(\mathcal L(D\mid\theta)\equiv P(D\mid\theta)\) is the likelihood
function.\footnote{The distinction between the likelihood
\({\mathcal L}(D\mid\theta)\equiv P(D\mid\theta)\) and the posterior
\(P(\theta\mid D)\) is important.  The quantity \(P(D\mid\theta)\) is
the conditional probability density of the data \(D\) for a fixed value
of the parameter \(\theta\).  In that interpretation, \(\theta\) is
assumed to be given, while the data \(D\) are regarded as the random
variable.  Once the experiment has been performed, however, the observed
dataset \(D\) is fixed.  The same expression \(P(D\mid\theta)\), now
viewed as a function of \(\theta\) for this fixed dataset, is called the
likelihood and is denoted by \({\mathcal L}(D\mid\theta)\).  Thus, the
likelihood is not a probability density in \(\theta\); rather, it
measures how well different values of \(\theta\) account for the
observed data.  The posterior distribution is instead
\(P(\theta\mid D)\), which describes how plausible different values of
\(\theta\) are after the data have been taken into account.}
The normalization factor
\begin{align}
P(D)=\int d\theta\;\mathcal L(D\mid\theta)P(\theta)
\label{eq:evidence_theta}
\end{align}
is the Bayesian evidence.  For more details of the Bayesian formalism,
see Ref.~\cite{Muk16OBa}.

In the present analysis, the role of \(\theta\) is played by the
coefficient vector
\begin{align}
\bm a=(a_0,a_1,a_2),
\label{a1}
\end{align}
which specifies the quadratic representation of the smooth function
\(y(E)=\log_{10}S^{*}(E)\). The posterior for the coefficient vector is
therefore
\begin{align}
P(\bm a\mid D)=\frac{P(D\mid\bm a)\,P(\bm a)}{P(D)}.
\label{eqbayes_a_general}
\end{align}

In the present notation, the likelihood is denoted by
\begin{align}
\mathcal L_{\mathrm{global}}(D\mid\bm a)\equiv P(D\mid\bm a),
\end{align}
and the prior by \(P(\bm a)\).

Since the parameter vector \(\bm a\) is continuous, the normalization
factor is given by
\begin{align}
P(D)=\int d\bm a\;
\mathcal L_{\mathrm{global}}(D\mid\bm a)\,P(\bm a).
\label{eqevidence}
\end{align}
Accordingly, the posterior can be written explicitly as
\begin{align}
P(\bm a\mid D)=
\frac{\mathcal L_{\mathrm{global}}(D\mid\bm a)\,P(\bm a)}
{\int d\bm a\;
\mathcal L_{\mathrm{global}}(D\mid\bm a)\,P(\bm a)}.
\label{eqbayes_a_coeff}
\end{align}
In the present analysis, a flat prior is adopted for the coefficient
vector \(\bm a\), i.e.
\[
P(\bm a)=\mathrm{const}.
\]
Hence,
\begin{align}
P(\bm a\mid D)\propto
\mathcal L_{\mathrm{global}}(D\mid\bm a).
\label{eqbayes_c}
\end{align}
Accordingly, the MAP (maximum a posteriori) coefficient vector is attained
at the same point as the maximum-likelihood estimate, and therefore at
the same point as the minimum-\(\chi^2\) solution.

The dataset \(D\) consists of the adopted direct measurements and the
inverse-kinematics data of Ref.~\cite{NanWang}, each characterized by a
measured or derived value of the modified astrophysical factor
\(S^{*}\) and an associated uncertainty. The aim is not to infer
\(S^{*}(E)\) independently at each energy, but rather to determine a
smooth global representation that is statistically consistent with the
adopted dataset and suitable for extrapolation to lower energies.

The model relation between the coefficient vector and the modified
astrophysical factor is
\begin{align}
S^{*}(E;\bm a)=10^{\,y(E;\bm a)}
          =10^{\,a_0+a_1(E-E_0)+a_2(E-E_0)^2}.
\label{SstarEa1}
\end{align}

For the fixed observed dataset \(D\), the posterior \(P(\bm a\mid D)\)
does not select a single coefficient vector. Instead, it assigns
probabilities to many possible coefficient vectors \(\bm a\) that are
compatible with the data. Each such coefficient vector defines one
smooth curve \(S^{*}(E;\bm a)\). Thus, the posterior \(P(\bm a\mid D)\)
induces not one single curve, but an ensemble of curves consistent with
the observed dataset \(D\).

This point is important. The posterior is obtained first in the space of
the polynomial coefficients \(\bm a\), not directly in the space of the
modified astrophysical factor \(S^{*}(E)\). The role of
\(P(\bm a\mid D)\) is to tell us which coefficient vectors $\bm a$  are more
probable and which are less probable after the data have been taken into
account. Once a particular coefficient vector \(\bm a\) is chosen,
Eq.~(\ref{SstarEa1}) gives the corresponding value of the modified
astrophysical factor at every energy \(E\), that is, it gives one
definite smooth curve \(S^{*}(E;\bm a)\).

At any fixed energy \(E\), the ensemble of allowed curves gives a
corresponding ensemble of values \(S^{*}(E;\bm a)\). In this way, the
posterior \(P(\bm a\mid D)\) in coefficient space induces a posterior
distribution for the modified astrophysical factor at that same energy,
denoted by \(P(S^{*}(E)\mid D)\). In other words, the posterior
distribution for \(S^{*}(E)\) is obtained by evaluating
\(S^{*}(E;\bm a)\) for all coefficient vectors allowed by the posterior
\(P(\bm a\mid D)\).

The way this is implemented in practice is through posterior sampling.
Regions of coefficient space where \(P(\bm a\mid D)\) is large are
sampled densely when evaluating Eq.~(\ref{SstarEa1}), whereas regions
where \(P(\bm a\mid D)\) is small are sampled sparsely. Therefore, the
posterior \(P(\bm a\mid D)\) determines how the sampled values
\(S^{*}(E;\bm a)\) are distributed at each fixed energy \(E\).
Coefficient vectors $\bm a$ that are strongly supported by the data contribute
many sampled values of \(S^{*}(E;\bm a)\), while coefficient vectors
that are weakly supported contribute few.

The central \(68\%\) credible interval for \(S^{*}(E)\) is then obtained
from the distribution of these sampled values \(S^{*}(E;\bm a)\) by
determining their 16th and 84th percentiles. Specifically,
\(S^{*,16\%}(E)\) is defined so that \(16\%\) of the sampled values of
\(S^{*}(E;\bm a)\) lie below it, while \(S^{*,84\%}(E)\) is defined so
that \(84\%\) of the sampled values lie below it. The interval
\begin{align}
\left[S^{*,16\%}(E),\,S^{*,84\%}(E)\right]
\end{align}
therefore gives the central \(68\%\) credible interval for the modified
astrophysical factor at the chosen energy.

It is useful to emphasize that the distribution in coefficient space and
the distribution in \(S^{*}\)-space are not the same object. The
posterior \(P(\bm a\mid D)\) describes the allowed values of the
coefficient vector \(\bm a\), whereas \(P(S^{*}(E)\mid D)\) describes
the allowed values of the modified astrophysical factor at fixed energy.
These two distributions do not, in general, coincide, because the
mapping \(\bm a \mapsto S^{*}(E;\bm a)\) is nonlinear.

In particular, the MAP coefficient vector \(\bm a_{\rm MAP}\) defines
the MAP curve
\begin{align}
S^{*,{\rm MAP}}(E)=S^{*}(E;\bm a_{\rm MAP}),
\end{align}
but the central \(68\%\) region in coefficient space does not, in
general, coincide with the central \(68\%\) interval of \(S^{*}(E)\).
Likewise, the percentile curves \(S^{*,16\%}(E)\) and
\(S^{*,84\%}(E)\) are constructed from the distribution of the sampled
values \(S^{*}(E;\bm a)\) at fixed energy and do not usually correspond
to one single coefficient vector.

Thus, \(P(\bm a\mid D)\) plays the role of an intermediate quantity: it
encodes which coefficient values are consistent with the data and their
uncertainties, and, through the nonlinear mapping \(S^{*}(E;\bm a)\),
induces the posterior distribution, median trend, MAP curve, and
credible interval for the modified astrophysical factor. The overall
workflow of the analysis is therefore
\begin{align}
D \;\longrightarrow\; \mathcal L_{\mathrm{global}}(D\mid\bm a)
\;\longrightarrow\; P(\bm a\mid D)
\;\longrightarrow\; S^{*}(E;\bm a).
\label{workflow}
\end{align}

\section{Global Posterior Construction}
\label{sec:globalposterior}

The Bayesian inference is performed in logarithmic space by defining
\begin{align}
y(E)\equiv \log_{10}S^{*}(E),
\label{eqydef}
\end{align}
where the modified astrophysical factor is
\begin{align}
S^{*}(E)=S(E)\,\exp(0.46\,E),
\label{eqSstar}
\end{align}
with \(E\) expressed in MeV.  This modified \(S\) factor is commonly
used in analyses of the \(^{12}{\rm C}+{}^{12}{\rm C}\) fusion reaction
\cite{Patterson}.  The smooth trend of \(y(E)\) is represented by the
quadratic parametrization of Eq.~(\ref{eqpoly1}).

\subsection{Global likelihood in log space}

Each experimental point is mapped to the logarithmic variable and
represented by \((E_i,y_i,\sigma_{y,i})\), where \(\sigma_{y,i}\) is the
uncertainty assigned in log space.  The log-space uncertainty is obtained
from the effective relative uncertainty by standard error propagation,
\begin{align}
\sigma_{y,i}^{\rm like}=\frac{\delta_i^{\rm eff}}{\ln 10}.
\label{eqsigma_log_like_postfig}
\end{align}
Here \(\delta_i^{\rm eff}\) includes the quoted point-to-point
uncertainty and the conservative fractional systematic component
discussed in Sec.~\ref{secdata}.

Assuming independent Gaussian uncertainties in \(y\), each experimental
point is written as
\begin{align}
y_i = y(E_i;\bm a)+\epsilon_i,
\label{eq:yi_model_postfig}
\end{align}
where \(\epsilon_i\) is the deviation of the measured value from the
model prediction at energy \(E_i\).  The random variable \(\epsilon_i\)
is assumed to follow a normal distribution with zero mean and variance
\((\sigma_{y,i}^{\rm like})^2\),
\begin{equation}
f(\epsilon_i)=
\frac{1}{\sqrt{2\pi}\,\sigma_{y,i}^{\rm like}}
\exp\!\left[-\frac{\epsilon_i^2}{2(\sigma_{y,i}^{\rm like})^2}\right].
\label{eq:eps_density_postfig}
\end{equation}
Since
\[
\epsilon_i=y_i-y(E_i;\bm a),
\]
the global likelihood is
\begin{align}
\mathcal{L}_{\rm global}(D\mid\bm a)
&=
\prod_{i=1}^{N}
\frac{1}{\sqrt{2\pi}\,\sigma_{y,i}^{\rm like}}
\exp\!\left[
-\frac{\bigl(y_i-y(E_i;\bm a)\bigr)^2}
{2\,(\sigma_{y,i}^{\rm like})^2}
\right],
\label{eq:global_like_compact_postfig}
\end{align}
where \(N\) is the total number of points in the adopted dataset.

In \(\mathcal{L}_{\rm global}(D\mid\bm a)\), the observed dataset \(D\)
is fixed, while the coefficient vector \(\bm a\) is varied.  Thus
\(\mathcal{L}_{\rm global}(D\mid\bm a)\equiv P(D\mid\bm a)\) is the
likelihood, not the posterior.  The posterior is obtained only after
combining the likelihood with the prior through Bayes' theorem,
\begin{align}
P(\bm a\mid D)=
\frac{\mathcal{L}_{\rm global}(D\mid\bm a)\,P(\bm a)}
{\int d\bm a\;\mathcal{L}_{\rm global}(D\mid\bm a)\,P(\bm a)}.
\label{eqbayes_a_like_postfig}
\end{align}
For Gaussian experimental uncertainties and a flat prior,
\[
P(\bm a)=\mathrm{const},
\]
the posterior is proportional to the likelihood.  Therefore, the most
probable coefficient vector, denoted by
\(\bm a_{\rm MAP}\) with MAP standing for maximum a posteriori, is the
value of \(\bm a\) that maximizes the posterior probability:
\begin{align}
\bm a_{\rm MAP}
=
\operatorname*{arg\,max}_{\bm a}\,P(\bm a\mid D).
\label{eqaMAP_postfig}
\end{align}
For Gaussian experimental uncertainties and a flat prior,
\[
P(\bm a)=\mathrm{const},
\]
the posterior is proportional to the likelihood.  Therefore, the most
probable coefficient vector, denoted by
\(\bm a_{\rm MAP}\) with MAP standing for maximum a posteriori, is the
value of \(\bm a\) that maximizes the posterior probability:
\begin{align}
\bm a_{\rm MAP}
=
\operatorname*{arg\,max}_{\bm a}\,P(\bm a\mid D).
\label{eqaMAP_postfig}
\end{align}
For the Gaussian likelihood used here, maximizing the posterior is
equivalent to minimizing \(\chi^2(\bm a)\), where
\[
\chi^2(\bm a)=
\sum_{i=1}^{N}
\frac{\epsilon_i^2}{\sigma_{y,i}^2},
\qquad
\epsilon_i=y_i-y(E_i;\bm a).
\]
Thus, for the flat prior adopted here, \(\bm a_{\rm MAP}\) coincides
with the minimum-\(\chi^2\) solution.

\subsection{Posterior of the polynomial coefficients}
\label{Posteriorpolyncoeff1_postfig}

For the quadratic representation of \(\log_{10}S^{*}(E)\), the unknown
coefficients are
\[
\bm a=(a_0,a_1,a_2)^{\mathsf T}.
\]
The data do not fix these coefficients exactly.  Rather, they allow a
range of nearby values around the best-fit point \(\bm a_{\rm MAP}\).
In the neighborhood of this point, the posterior distribution can be
approximated by a three-dimensional Gaussian,
\begin{align}
P(\bm a\mid D)\approx
\mathcal N\!\left(\bm a_{\rm MAP},\mathbf C_{\bm a}\right),
\label{eqnormal_expl_short_postfig}
\end{align}
or explicitly,
\begin{align}
P(\bm a\mid D)
&\approx
\frac{1}{(2\pi)^{3/2}|\mathbf C_{\bm a}|^{1/2}}
\nonumber\\
&\times
\exp\!\left[
-\frac12
(\bm a-\bm a_{\rm MAP})^{\mathsf T}
\mathbf C_{\bm a}^{-1}
(\bm a-\bm a_{\rm MAP})
\right].
\label{eqnormal_expl_postfig}
\end{align}
The covariance matrix \(\mathbf C_{\bm a}\) determines the uncertainties
and correlations of the fitted coefficients.  Technical details of
\(\mathbf C_{\bm a}\) are given in Appendix~\ref{Covariance_matrix}.

This local Gaussian approximation is appropriate because the model is
linear in the coefficients and the uncertainties are treated as Gaussian
in \(\log_{10}S^{*}(E)\).  The posterior is also constrained by a large
number of experimental points over a broad energy interval.

\subsection{Uncertainty propagation to \(S^{*}(E)\)}
\label{subsec:uncertainty_propagation}

To propagate the uncertainties of the fitted coefficients to derived
quantities, the Gaussian posterior of Eq.~(\ref{eqnormal_expl}) is
sampled by Monte Carlo.  Each realization of the coefficient vector is
generated as
\begin{align}
\bm a^{(k)}=\bm a_{\rm MAP}
+\sum_{i=1}^{3}\sqrt{\lambda_i}\,\xi_i^{(k)}\,\bm e_i,
\label{eqMCsampling_simple}
\end{align}
where \(\lambda_i\) and \(\bm e_i\) are the eigenvalues and eigenvectors
of \(\mathbf C_{\bm a}\), and \(\xi_i^{(k)}\) are independent standard
normal random variables.  Each sampled coefficient vector defines one
smooth curve,
\begin{align}
S^{*}(E;\bm a^{(k)})=10^{\,y(E;\bm a^{(k)})}.
\label{eqSstar_sample_curve}
\end{align}

The Monte Carlo procedure samples only the coefficient posterior
\(P(\bm a\mid D)\); no experimental data are resampled or modified at
this stage.  The resulting ensemble of curves is then used to determine
the posterior median and the corresponding \(68\%\) credible interval
for the smooth component \(S^{*}(E)\).

At a fixed energy \(E\), the sampled coefficient vectors generate a
corresponding set of sampled values \(S^{*}(E;\bm a^{(k)})\).  The
16th, 50th, and 84th percentiles of this distribution define,
respectively, the lower limit, median, and upper limit of the central
\(68\%\) credible interval.  Thus the posterior of the polynomial
coefficients induces a posterior distribution for \(S^{*}(E)\) at every
energy.

It is important to distinguish the coefficient posterior from the
posterior distribution of \(S^{*}(E)\).  The posterior
\(P(\bm a\mid D)\) describes the allowed values of the coefficient
vector, while the distribution of \(S^{*}(E;\bm a)\) describes the
allowed values of the smooth modified astrophysical factor at fixed
energy.  Because the mapping
\[
\bm a \mapsto S^{*}(E;\bm a)
\]
is nonlinear, the MAP curve, the median curve, and the percentile curves
need not correspond to the same coefficient vector.

\begin{figure*}[t]
  \centering
  \includegraphics[width=0.80\textwidth]{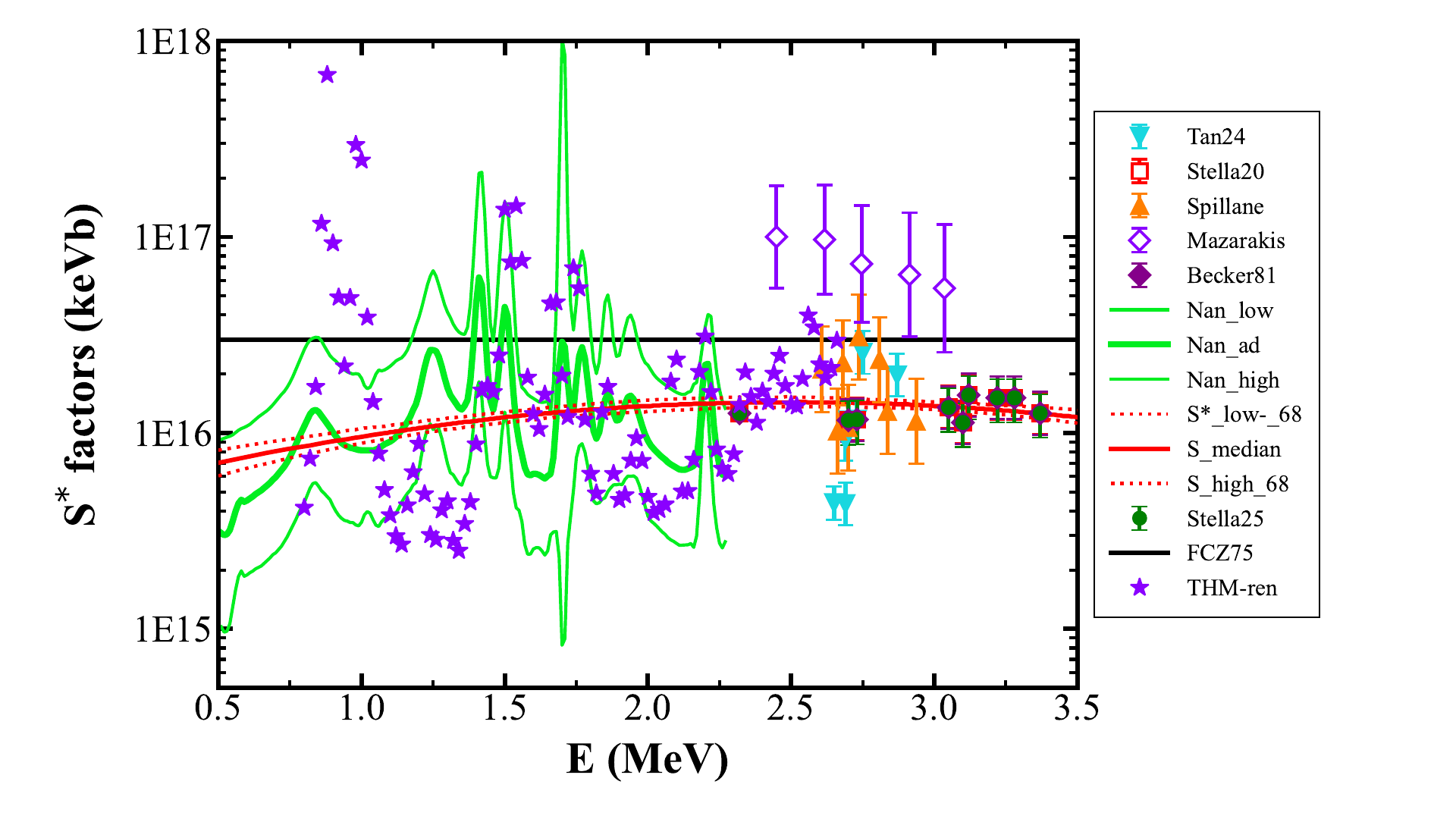}
  \caption{Posterior smooth extrapolation of \(S^{*}(E)\) for the
  \(^{12}{\rm C}+{}^{12}{\rm C}\) fusion reaction.  The solid red curve
  shows the posterior median of the adopted Bayesian analysis, while the
  red dotted curves show the corresponding central \(68\%\) credible
  interval.  The posterior is constructed from the direct measurements
  and the inverse-kinematics data of Ref.~\cite{NanWang}.  The
  Coulomb--nuclear-renormalized THM points are shown for comparison, but
  are not included as an independent constraint in the likelihood used to
  construct the posterior.  The horizontal black line denotes the
  Fowler--Caughlan--Zimmerman reference value
  \(S^{*}_{\rm FCZ75}=3.0\times10^{16}~{\rm keV\,b}\)
  \cite{FCZ75}.}
  \label{figpost_linS}
\end{figure*}

The resulting global posterior in linear space is shown in
Fig.~\ref{figpost_linS}.  The solid red curve represents the posterior
median of the adopted smooth global component, while the red dotted
curves show the corresponding central \(68\%\) credible interval.  The
Coulomb--nuclear-renormalized THM data are shown for comparison, but are
not included as an independent constraint in the likelihood used to
construct the posterior.  The plotted posterior should therefore be
interpreted as the smooth global trend constrained by the adopted
dataset, not as a resonance-by-resonance description of the low-energy
fusion cross section.

\subsection{Interpretation of the global posterior}
\label{subsec:global_posterior_interpretation}

The posterior band in Fig.~\ref{figpost_linS} represents the uncertainty
of the smooth global component of \(S^{*}(E)\), not the full scatter of
the experimental data.  The adopted data contain local resonance
structures and point-to-point fluctuations.  These structures enter the
likelihood through the measured values of \(S^{*}(E)\) and therefore
influence the fitted normalization, slope, and curvature of the smooth
posterior.  The quadratic regression does not, however, assign separate
resonance energies, widths, or interference phases to individual
structures.  Thus, the fitted curve should be viewed as a smooth average
constrained by resonance-affected data rather than as a
resonance-by-resonance parametrization.

A qualitative comparison of the resonance-like structures in the
Coulomb--nuclear-renormalized THM extraction \cite{MukTHM}  with those in the
inverse-kinematics data of Ref.~\cite{NanWang} supports treating the THM
data as an external comparison rather than as an independent likelihood
constraint.  The energies quoted here are used only to indicate the
proximity of the structures and should not be interpreted as precise
resonance centroids.  The low-energy THM structure near
\(E\simeq0.88~{\rm MeV}\) has a nearby counterpart in the
inverse-kinematics data near \(E\simeq0.83~{\rm MeV}\), although the THM
strength is larger.  Conversely, the structures near \(E\simeq1.24\)
and \(1.42~{\rm MeV}\) in Ref.~\cite{NanWang} do not have clear
counterparts in the renormalized THM extraction.  At higher energies,
several resonance-like structures occur at similar energies in the two
datasets, for example near \(E\simeq1.50\), \(1.66\)--\(1.69\),
\(1.74\)--\(1.77\), \(1.85\), \(1.94\)--\(1.96\), and
\(2.20~{\rm MeV}\).  In these cases the THM structures are generally
weaker than those in Ref.~\cite{NanWang}.  Overall, the
Coulomb--nuclear-renormalized THM points tend to lie near the lower edge
of the inverse-kinematics band of Ref.~\cite{NanWang}, rather than
defining an independent higher normalization of the low-energy
\(S^{*}(E)\).  The THM feature near \(E\simeq2.09~{\rm MeV}\) does not
have a clear counterpart in Ref.~\cite{NanWang}.

This comparison shows that the renormalized THM extraction contains
resonance-like information that significantly overlaps with the
inverse-kinematics constraint.  At the same time, the absolute strengths
of the renormalized THM structures retain some dependence on the
reaction-model assumptions entering the THM extraction.  Including both
datasets as independent constraints with comparable statistical weight
would therefore risk double counting part of the same low-energy
resonance information.  Including both datasets as
independent constraints with comparable statistical weight would
therefore risk double counting part of the same low-energy structure.
For this reason, the inverse-kinematics data of Ref.~\cite{NanWang} are
included in the likelihood, whereas the Coulomb--nuclear-renormalized
THM data are shown as an external comparison.

The comparison of the Coulomb--nuclear-renormalized THM extraction \cite{MukTHM}  with those in the
inverse-kinematics data of Ref.~\cite{NanWang} is nevertheless important.  The original plane-wave THM analysis \cite{Tumino} produced a very large low-energy enhancement of the
\(^{12}{\rm C}+{}^{12}{\rm C}\) astrophysical factor, which is not
supported by more recent direct and inverse-kinematics measurements
\cite{Tan20,Tan24,Fruet,Nippert,NanWang} or by modern theoretical
calculations~\cite{Kimura21,Kimura23,Descouvemont26}.  After inclusion
of Coulomb and nuclear distortion effects, the renormalized THM strength
is strongly reduced and becomes much closer to the modern low-energy
constraints than the original plane-wave THM extraction.

As a compact diagnostic of the smooth extrapolated component, the
posterior at \(E=1.5~{\rm MeV}\) gives
\begin{align}
S_{\rm global}^{*}(1.5~{\rm MeV})
=
\left(
1.13,\;1.20,\;1.28
\right)\times10^{16}\ {\rm keV\,b},
\label{eq:S15_result}
\end{align}
where the three values denote the 16th, 50th, and 84th percentiles,
respectively.  The quantity \(S_{\rm global}^{*}(1.5~{\rm MeV})\)
should not be interpreted as a resonance-resolved observable.  It is
used only as a compact diagnostic of the adopted smooth global component
at the reference energy \(E=1.5~{\rm MeV}\).

The median value obtained here is noticeably lower than the
Fowler--Caughlan--Zimmerman reference normalization
\[
S^{*}_{\rm FCZ75}=3.0\times10^{16}\ {\rm keV\,b}
\]
\cite{FCZ75}.  This comparison shows that the updated posterior does not
support the traditional FCZ75-like normalization of the smooth
low-energy trend.

The astrophysically relevant quantity is the thermonuclear reaction
rate.  In the present work this rate is constructed by retaining the
inverse-kinematics input of Ref.~\cite{NanWang} in the low-energy region,
where the measured resonance structures enter explicitly, and by
matching this input to the smooth global posterior at higher energies.
\subsection{Combination of heterogeneous datasets}

The Bayesian framework provides a transparent way to combine datasets
with different normalizations, energy coverages, and quoted
uncertainties within a single global analysis.  All datasets included in
the adopted fit enter the same likelihood, and their relative influence
is determined by the assigned uncertainties rather than by arbitrary
manual weighting.

Because the purpose of the present global fit is to extract the smooth
trend of \(S^{*}(E)\), rather than to reproduce every localized
structure in the data, the fit is performed in the logarithmic variable
\[
y(E)=\log_{10}S^{*}(E),
\]
with a conservative uncertainty assignment.  Working in log space
stabilizes the global fit and reduces the dynamic range of the data.
The final posterior is then transformed back to linear \(S^{*}(E)\)
space for physical interpretation and for reaction-rate calculations.

This procedure reduces the risk that one dense subset of points or one
localized structure dominates the global trend.  Consequently, the
inferred posterior reflects the common smooth component supported by the
adopted dataset, while still retaining the overall experimental
information carried by the individual measurements.

\subsection{Interpretation of the posterior width}

When many datasets with different normalizations and quoted
uncertainties are constrained to a single smooth global trend, only a
limited range of background curves can provide an acceptable overall
description.  The resulting posterior band is therefore narrow not
because the fit attempts to reproduce fine resonance structure, but
because the smooth component is constrained simultaneously by a large
body of data.

The width of the posterior should be interpreted as the uncertainty of
the smooth global trend within the adopted likelihood and uncertainty
model.  It is not a measure of the amplitude of local resonant
fluctuations.  Localized resonance structures remain present in the
data and influence the weighted global fit, but they are not modeled
individually in the quadratic regression.

Thus, the posterior band is meaningful for two purposes: it quantifies
the uncertainty of the smooth extrapolated component, and it provides
the input for the thermonuclear reaction-rate calculation.  The
inverse-kinematics data of Ref.~\cite{NanWang} provide the direct
low-energy constraint included in the likelihood, while the
Coulomb--nuclear-renormalized THM data are shown only for comparison.

\section{Thermonuclear reaction rate}

The combined reaction rate is constructed by using the direct
inverse-kinematics input of Ref.~\cite{NanWang} in the low-energy region
and matching it to the global smooth posterior at higher energies.
Hereafter, the input of Ref.~\cite{NanWang} is referred to as the
inverse-kinematics data.  These data dominate the rate at the lowest temperatures, where the
Gamow window lies largely within the measured sub-Coulomb energy region.
As the temperature increases, the Gamow window shifts to higher
energies, and the contribution from the global smooth posterior above
the inverse-kinematics interval becomes increasingly important.  
Hereafter, we refer to the input of Ref.~\cite{NanWang} as the
inverse-kinematics data.  These data dominate the rate at the lowest
temperatures, where the Gamow window lies largely within the measured
sub-Coulomb energy region.  As the temperature increases, the Gamow
window shifts to higher energies, and the contribution from the smooth
global posterior above the inverse-kinematics interval becomes
increasingly important.

For hydrostatic carbon burning at \(T_9\lesssim 0.7\), the effective
energy window lies mainly below \(2~{\rm MeV}\).  In this temperature
range the rate is controlled primarily by the low-energy behavior of
\(S^{*}(E)\), and the direct inverse-kinematics constraints are therefore
especially important.  At higher temperatures, \(T_9\gtrsim 1\), the
Gamow window shifts to higher energies and the rate becomes increasingly
sensitive to the smooth global posterior above \(2~{\rm MeV}\).

Because the present numerical integration is restricted to the interval
\[
1.0 \le E \le 3.5~{\rm MeV},
\]
the resulting reaction rate should be regarded as complete only when the
dominant Gamow window is contained within this energy range.  For the
\(^{12}{\rm C}+{}^{12}{\rm C}\) system, the Gamow peak may be estimated
as
\[
E_0 \simeq 2.42\,T_9^{2/3}~{\rm MeV},
\]
with an approximate width
\[
\Delta E \simeq 1.05\,T_9^{5/6}~{\rm MeV}.
\]
Thus the upper part of the effective energy window is approximately
\[
E_{\rm upper}\simeq E_0+\frac{\Delta E}{2}.
\]
Representative values are shown in Table~\ref{tab:gamow_window}.

\begin{table}[t]
\centering
\caption{Approximate Gamow peak energy \(E_0\) and upper edge
\(E_{\rm upper}\simeq E_0+\Delta E/2\) for
\(^{12}{\rm C}+{}^{12}{\rm C}\).  Energies are given in MeV.}
\label{tab:gamow_window}
\begin{tabular}{ccc}
\hline
\(T_9\) & \(E_0\) & \(E_{\rm upper}\) \\
\hline
0.5 & 1.52 & 1.82 \\
0.7 & 1.91 & 2.30 \\
1.0 & 2.42 & 2.95 \\
1.2 & 2.73 & 3.34 \\
1.3 & 2.88 & 3.54 \\
\hline
\end{tabular}
\end{table}

Consequently, the present rate is most reliable for
\[
T_9 \lesssim 1.2\text{--}1.3,
\]
where the effective energy window remains essentially inside the adopted
integration interval.  At higher temperatures, and especially near
\(T_9\simeq 2\), the Gamow window extends significantly above
\(3.5~{\rm MeV}\).  In that regime, contributions from
\(E>3.5~{\rm MeV}\) become important, and the calculated rate should be
viewed as a truncated rate rather than a complete stellar reaction rate.

The numerical values of the present reaction rate, the CF88 analytic
rate, and the corresponding median ratio \(R_{\rm present}/R_{\rm CF88}\)
are listed in Table~\ref{tab:present_c12c12_rate}.

\begin{table*}[t]
\centering
\scriptsize
\setlength{\tabcolsep}{3pt}
\renewcommand{\arraystretch}{0.90}
\caption{\(^{12}{\rm C}+{}^{12}{\rm C}\) thermonuclear reaction rate
obtained from the updated posterior. The columns labeled low, median,
and high correspond to the 16th, 50th, and 84th percentiles of the
posterior rate distribution. The CF88 analytic rate is shown for
comparison, together with the ratio of the present median rate to the
CF88 rate. Rates are given in units of
\({\rm cm^{3}\,mol^{-1}\,s^{-1}}\).}
\label{tab:present_c12c12_rate}
\begin{tabular}{cccccc}
\hline
\(T_9\) &
\(N_A\langle\sigma v\rangle_{\rm low}\) &
\(N_A\langle\sigma v\rangle_{\rm med}\) &
\(N_A\langle\sigma v\rangle_{\rm high}\) &
\(N_A\langle\sigma v\rangle_{\rm CF88}\) &
\(R_{\rm med}/R_{\rm CF88}\) \\
\hline
0.30 & \(4.425\times10^{-29}\) & \(4.756\times10^{-29}\) & \(5.170\times10^{-29}\) & \(1.437\times10^{-28}\) & 0.3311 \\
0.35 & \(2.185\times10^{-26}\) & \(2.336\times10^{-26}\) & \(2.519\times10^{-26}\) & \(6.662\times10^{-26}\) & 0.3507 \\
0.40 & \(3.623\times10^{-24}\) & \(3.862\times10^{-24}\) & \(4.142\times10^{-24}\) & \(1.048\times10^{-23}\) & 0.3687 \\
0.45 & \(2.708\times10^{-22}\) & \(2.882\times10^{-22}\) & \(3.081\times10^{-22}\) & \(7.485\times10^{-22}\) & 0.3850 \\
0.50 & \(1.106\times10^{-20}\) & \(1.176\times10^{-20}\) & \(1.256\times10^{-20}\) & \(2.943\times10^{-20}\) & 0.3996 \\
0.55 & \(2.816\times10^{-19}\) & \(2.995\times10^{-19}\) & \(3.197\times10^{-19}\) & \(7.261\times10^{-19}\) & 0.4125 \\
0.60 & \(4.918\times10^{-18}\) & \(5.232\times10^{-18}\) & \(5.585\times10^{-18}\) & \(1.234\times10^{-17}\) & 0.4238 \\
0.65 & \(6.314\times10^{-17}\) & \(6.717\times10^{-17}\) & \(7.173\times10^{-17}\) & \(1.549\times10^{-16}\) & 0.4336 \\
0.70 & \(6.281\times10^{-16}\) & \(6.682\times10^{-16}\) & \(7.134\times10^{-16}\) & \(1.512\times10^{-15}\) & 0.4420 \\
0.75 & \(5.043\times10^{-15}\) & \(5.362\times10^{-15}\) & \(5.723\times10^{-15}\) & \(1.195\times10^{-14}\) & 0.4489 \\
0.80 & \(3.373\times10^{-14}\) & \(3.584\times10^{-14}\) & \(3.823\times10^{-14}\) & \(7.887\times10^{-14}\) & 0.4544 \\
0.85 & \(1.928\times10^{-13}\) & \(2.047\times10^{-13}\) & \(2.181\times10^{-13}\) & \(4.463\times10^{-13}\) & 0.4586 \\
0.90 & \(9.614\times10^{-13}\) & \(1.019\times10^{-12}\) & \(1.085\times10^{-12}\) & \(2.209\times10^{-12}\) & 0.4615 \\
0.95 & \(4.251\times10^{-12}\) & \(4.502\times10^{-12}\) & \(4.785\times10^{-12}\) & \(9.724\times10^{-12}\) & 0.4630 \\
1.00 & \(1.690\times10^{-11}\) & \(1.788\times10^{-11}\) & \(1.897\times10^{-11}\) & \(3.862\times10^{-11}\) & 0.4629 \\
1.10 & \(2.025\times10^{-10}\) & \(2.138\times10^{-10}\) & \(2.264\times10^{-10}\) & \(4.677\times10^{-10}\) & 0.4571 \\
1.20 & \(1.774\times10^{-9}\) & \(1.870\times10^{-9}\) & \(1.977\times10^{-9}\) & \(4.224\times10^{-9}\) & 0.4426 \\
1.30 & \(1.195\times10^{-8}\) & \(1.258\times10^{-8}\) & \(1.329\times10^{-8}\) & \(3.004\times10^{-8}\) & 0.4188 \\
1.40 & \(6.436\times10^{-8}\) & \(6.774\times10^{-8}\) & \(7.159\times10^{-8}\) & \(1.752\times10^{-7}\) & 0.3865 \\
1.50 & \(2.865\times10^{-7}\) & \(3.015\times10^{-7}\) & \(3.188\times10^{-7}\) & \(8.659\times10^{-7}\) & 0.3482 \\
1.60 & \(1.083\times10^{-6}\) & \(1.140\times10^{-6}\) & \(1.206\times10^{-6}\) & \(3.715\times10^{-6}\) & 0.3068 \\
1.70 & \(3.553\times10^{-6}\) & \(3.741\times10^{-6}\) & \(3.960\times10^{-6}\) & \(1.411\times10^{-5}\) & 0.2651 \\
1.80 & \(1.032\times10^{-5}\) & \(1.087\times10^{-5}\) & \(1.152\times10^{-5}\) & \(4.827\times10^{-5}\) & 0.2253 \\
1.90 & \(2.700\times10^{-5}\) & \(2.844\times10^{-5}\) & \(3.016\times10^{-5}\) & \(1.506\times10^{-4}\) & 0.1889 \\
2.00 & \(6.442\times10^{-5}\) & \(6.789\times10^{-5}\) & \(7.204\times10^{-5}\) & \(4.331\times10^{-4}\) & 0.1567 \\
\hline
\end{tabular}
\end{table*}

The reaction rate is obtained from a hybrid \(S^{*}(E)\) input.  For
\(E\lesssim2.2~{\rm MeV}\), the inverse-kinematics data of
Ref.~\cite{NanWang} are used so that the measured low-energy resonance
structures enter the Maxwellian rate integral explicitly.  For
\(E\gtrsim2.2~{\rm MeV}\), the adopted smooth posterior for
\(S^{*}(E)\) provides the higher-energy continuation.  The thermonuclear rate,
however, is determined by a Gamow-window average of the cross section
rather than by the local value of \(S^{*}(E)\) at a single energy.
Therefore, resonance structures affect the rate through their weighted
contribution over the effective energy window.

\begin{figure}[t]
\centering
\includegraphics[width=\columnwidth]{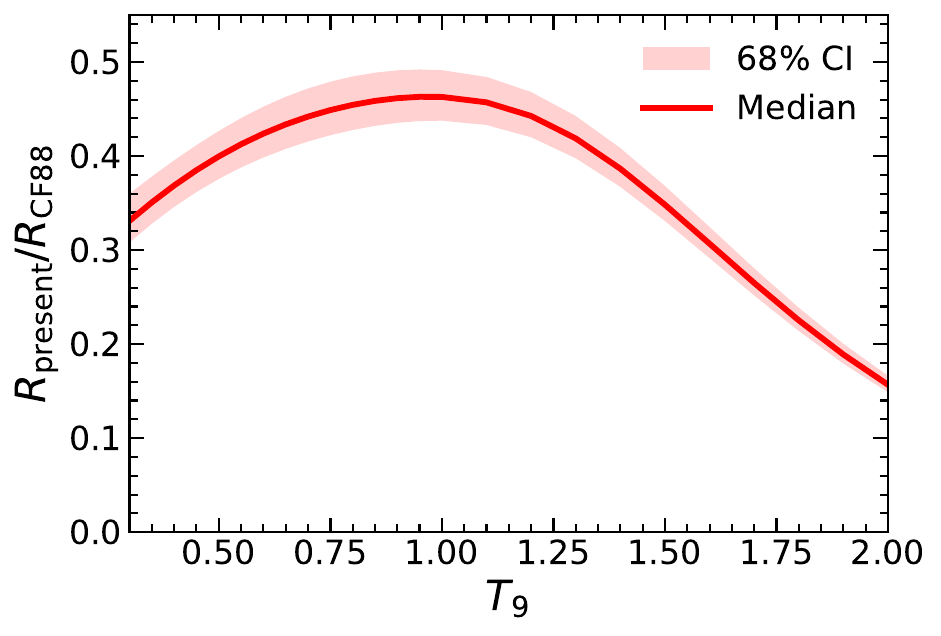}
\caption{Ratio of the present \(^{12}{\rm C}+{}^{12}{\rm C}\)
thermonuclear reaction rate to the CF88 analytic rate~\cite{Caughlan}.
The solid red curve shows the posterior median ratio, while the shaded
red region denotes the central \(68\%\) credible interval.  Values below
unity indicate that the present rate is lower than the CF88 rate.}
\label{fig:rate_cf88_ratio}
\end{figure}

As shown in Fig.~\ref{fig:rate_cf88_ratio}, the updated median rate is
systematically lower than the CF88 analytic rate~\cite{Caughlan} over
the temperature interval considered.  The median ratio
\(R_{\rm present}/R_{\rm CF88}\) remains below unity throughout this
range: it rises from about \(0.33\) at \(T_9=0.3\) to a maximum of about
\(0.46\) near \(T_9\simeq0.95\)--\(1.0\), and then decreases to
\(R_{\rm present}/R_{\rm CF88}\simeq0.16\) at \(T_9=2.0\).
This reduction reflects the lower smooth \(S^{*}(E)\) trend favored by
the posterior, as shown in Fig.~\ref{figpost_linS}, relative to
the CF88/FCZ75-like normalization~\cite{Caughlan,FCZ75}.  In particular,
the present median posterior gives
\[
S_{\rm global}^{*}(1.5~{\rm MeV})
=
1.20\times 10^{16}\ {\rm keV\,b},
\]
whereas the CF88 value is
\[
S_{\rm CF88}^{*}(1.5~{\rm MeV})
=
3.0\times 10^{16}\ {\rm keV\,b}.
\]
Thus, both the posterior \(S^{*}(E)\) and the thermonuclear reaction
rate indicate a lower normalization than the CF88/FCZ75-like trend.

The Coulomb--nuclear-renormalized THM data of Ref.~\cite{MukTHM} are
consistent with this downward shift in the sense that they do not
support the large low-energy enhancement of the original plane-wave THM
extraction.  However, as discussed above, they are shown only as an
external comparison and are not included as an independent constraint in
the likelihood used to construct the posterior.

\section{Conclusions}

A Bayesian analysis of the modified astrophysical factor
\(S^{*}(E)\) for the \(^{12}{\rm C}+{}^{12}{\rm C}\) fusion reaction has
been performed using a smooth quadratic parametrization of
\(\log_{10}S^{*}(E)\).  The adopted likelihood combines the available
direct measurements with the inverse-kinematics data of
Ref.~\cite{NanWang}.  The Coulomb--nuclear-renormalized THM data are
shown as an external comparison, but are not included as an independent
constraint in the likelihood, in order to avoid double counting of
overlapping low-energy resonance information and to reflect the residual
model dependence of the renormalized THM extraction.

The purpose of the global fit is not to provide a
resonance-by-resonance parametrization of the
\(^{12}{\rm C}+{}^{12}{\rm C}\) system, but rather to determine the
smooth global component of \(S^{*}(E)\) constrained by the adopted
heterogeneous data.  The adopted data contain local resonance
structures, and these structures enter the likelihood through the
measured values of \(S^{*}(E)\).  They therefore influence the fitted
normalization, slope, and curvature of the smooth posterior.  However,
the present fit does not assign separate resonance energies, widths, or
interference phases to individual structures.  The posterior curve
should therefore be interpreted as a smooth average constrained by
resonance-affected data, not as a local model of individual resonances.

A conservative uncertainty assignment was adopted to account for the
quoted point-to-point uncertainties together with residual
dataset-to-dataset normalization differences and other systematic
effects.  The fit was carried out in logarithmic space, where the large
dynamic range of \(S^{*}(E)\) is reduced and the experimental scatter is
more naturally represented by approximately Gaussian uncertainties.  The
posterior distribution of the polynomial coefficients was then
propagated to obtain the median smooth curve and the corresponding
central \(68\%\) credible interval for \(S^{*}(E)\).

As a compact diagnostic of the smooth extrapolated component, the global
posterior gives
\[
S_{\rm global}^{*}(1.5~{\rm MeV})
=
\left(
1.13,\;1.20,\;1.28
\right)\times10^{16}\ {\rm keV\,b},
\]
where the three values denote the 16th, 50th, and 84th percentiles,
respectively.  This quantity is used only as a benchmark for the smooth
global trend and should not be interpreted as a resonance-resolved
observable.  The median value is about \(40\%\) of the
Fowler--Caughlan--Zimmerman reference normalization
\[
S^{*}_{\rm FCZ75}=3.0\times10^{16}\ {\rm keV\,b}
\]
\cite{FCZ75}.

The main physical output of the analysis is the thermonuclear reaction
rate \(N_A\langle\sigma v\rangle\).  This rate is constructed from a
hybrid \(S^{*}(E)\) input.  In the low-energy region, the
inverse-kinematics data of Ref.~\cite{NanWang} are used so that the
measured resonance structures enter the Maxwellian rate integral
explicitly.  At higher energies, this input is matched to the Bayesian
smooth posterior.  Thus, the low-energy resonances observed in
Ref.~\cite{NanWang} enter the reaction rate directly, while the
higher-energy contribution is supplied by the statistically constrained
smooth global component.

The resulting median rate is systematically lower than the CF88 analytic
rate~\cite{Caughlan} over the temperature interval considered.  The
median ratio \(R_{\rm present}/R_{\rm CF88}\) remains below unity
throughout this range: it rises from about \(0.33\) at \(T_9=0.3\) to a
maximum of about \(0.46\) near \(T_9\simeq0.95\)--\(1.0\), and then
decreases to \(R_{\rm present}/R_{\rm CF88}\simeq0.16\) at
\(T_9=2.0\).  This behavior reflects the lower smooth \(S^{*}(E)\) trend
favored by the updated posterior relative to the traditional
FCZ75/CF88 normalization.

The present work does not rely on a specific hindrance model, nor does
it attempt a full \(R\)-matrix or resonance-by-resonance description.
Instead, it provides a transparent and reproducible Bayesian
determination of the smooth global component of \(S^{*}(E)\), together
with a thermonuclear reaction-rate estimate and an uncertainty band.
The obtained posterior lies in the direction of a reduced low-energy rate
relative to CF88, although it is obtained without imposing a
hindrance-model prior.  The framework can be updated straightforwardly
as new experimental or microscopic-theory constraints become available.

At temperatures for which the dominant Gamow window lies inside the
adopted energy interval, the calculated rate can be interpreted as the
rate implied by the adopted likelihood, uncertainty model, and hybrid
low-energy input.  At higher temperatures, where the Gamow window
extends above the upper end of the adopted energy interval, the rate
should be regarded as a truncated rate rather than a complete stellar
reaction rate.  Future direct measurements below \(E\simeq2~{\rm MeV}\),
as well as improved constraints on the low-energy resonance structure,
will be especially important for further reducing the uncertainty of the
\(^{12}{\rm C}+{}^{12}{\rm C}\) fusion rate under stellar conditions.

\section{Acknowledgments}
The author  thanks Alex Zhanov for technical assistance. 

\smallskip

\appendix

\section{Coveriance matrix}
\label{Covariance_matrix}
The covariance matrix \(\mathbf C_{\bm a}\) may be obtained from the
weighted regression or, equivalently, estimated from a Monte Carlo
sample of coefficient vectors \(\{\bm a^{(k)}\}\) through
\begin{align}
C_{ij}
=
\frac{1}{N_{\rm MC}-1}
\sum_{k=1}^{N_{\rm MC}}
\bigl(a_i^{(k)}-\bar a_i\bigr)
\bigl(a_j^{(k)}-\bar a_j\bigr),
\label{eq:app_samplecov}
\end{align}
where
\begin{align}
\bar a_i=
\frac{1}{N_{\rm MC}}
\sum_{k=1}^{N_{\rm MC}} a_i^{(k)}.
\label{eq:app_samplemean}
\end{align}
The matrix \(\mathbf C_{\bm a}\) determines the widths and correlations
of the allowed coefficient values and, through the mapping
\begin{align}
S^{*}(E)=10^{\,y(E)},
\end{align}
the posterior uncertainty band of the modified astrophysical factor.\section{Converting a log-space uncertainty into a fractional uncertainty}
\label{sec:logsigma_to_fraction}

In this work we often quantify uncertainties in log space,
\begin{equation}
y \equiv \log_{10} S,
\end{equation}
so that \(\sigma_y\) is the uncertainty in the variable \(y=\log_{10}S\) 

If $y$ is normally distributed
with standard deviation $\sigma_y$, then $S$ is log-normally distributed and the
$\pm 1\sigma$ band in $y$ corresponds to multiplicative factors in $S$:
\begin{equation}
S_{\pm} = 10^{\,y \pm \sigma_y} = S\,10^{\pm \sigma_y}.
\label{eq:mult_factor}
\end{equation}
Thus, the one-sigma \emph{multiplicative} factor is
\begin{equation}
f \equiv 10^{\sigma_y},
\end{equation}
and a convenient ``fractional'' one-sigma measure is
\begin{equation}
\frac{\Delta S}{S} \approx f-1 = 10^{\sigma_y}-1.
\label{eq:frac_from_dex}
\end{equation}
Conversely, a specified fractional uncertainty $\delta \equiv \Delta S/S$ corresponds to
\begin{equation}
\sigma_y = \log_{10}(1+\delta) \approx \frac{\delta}{\ln10}.
\label{eq:dex_from_frac}
\end{equation}

As a numerical example,
\begin{equation}
\sigma_y = 0.10
\quad\Rightarrow\quad
10^{0.10}= 1+\delta \simeq 1.2589 .
\end{equation}
Then
\begin{equation}
\frac{\Delta S}{S}\simeq \delta \simeq 0.259 \approx 26\%.
\end{equation}
Equivalently, a \(25\%\) fractional uncertainty corresponds to
\begin{equation}
\sigma_y = \log_{10}(1.25)\simeq 0.0969,
\end{equation}
which is very close to \(0.10\) in \(\log_{10}\) units.

\section{Chi-square, degrees of freedom, and the uncertainty calibration}
\label{app:chi2sigmasc}

Define the normalized residuals
\begin{equation}
r_i \equiv \frac{y_i-y(E_i;\bm a)}{\sigma_{y,i}},
\label{eq:ri_def}
\end{equation}
where \(y_i=\log_{10}S_i^{*}\) and \(\sigma_{y,i}\) is the assigned
uncertainty in log space.  The chi-square statistic and the reduced
chi-square are
\begin{equation}
\chi^2 \equiv \sum_{i=1}^{N} r_i^2,
\qquad
\chi_\nu^2 \equiv \frac{\chi^2}{\nu},
\qquad
\nu \equiv N-p,
\label{eq:chi2_defs}
\end{equation}
where \(N\) is the number of fitted data points, \(p\) is the number of
adjusted parameters, and \(\nu\) is the number of degrees of freedom.
In the present quadratic parametrization,
\[
p=3,
\]
since the three coefficients \((a_0,a_1,a_2)\) are fitted to the data.
Therefore,
\[
\nu=N-3.
\]

\subsection{Clipping of quoted uncertainties.}
Because the compiled dataset is heterogeneous, some points may carry unrealistically small or
excessively large quoted uncertainties in log space. To prevent a small subset of points from
dominating the global fit, I impose a conservative floor and cap,
\begin{align}
&\sigma_{y,i}\ \leftarrow\ \mathrm{Clip}\!\left(\sigma_{y,i};\,\sigma_{y,\min},\,\sigma_{y,\max}\right),
\nonumber\\
& \sigma_{y,\min}=0.10~\text{dex},\quad \sigma_{y,\max}=1.0~\text{dex},
\label{eq:sigmaclip_global}
\end{align}
and proceed with these clipped uncertainties as the adopted input.

\subsection{Clipping rule for log-space uncertainties.}
Each data point $y_i=\log_{10}S_i$ comes with an adopted log-space uncertainty $\sigma_{y,i}$
(in dex). To prevent a small subset of points with unrealistically small quoted errors from
dominating the weighted fit, and to avoid extremely large errors from effectively removing
points, I clip the uncertainties to a fixed interval:
\[
\sigma_{y,i}\leftarrow
\begin{cases}
\sigma_{y,\min}, & \sigma_{y,i}<\sigma_{y,\min},\\
\sigma_{y,i}, & \sigma_{y,\min}\le \sigma_{y,i}\le \sigma_{y,\max},\\
\sigma_{y,\max}, & \sigma_{y,i}>\sigma_{y,\max}.
\end{cases}
\]
In this work I use $\sigma_{y,\min}=0.10$ dex (approximately a $25\%$ relative uncertainty in $S$)
and $\sigma_{y,\max}=1.0$ dex.

\subsection{Additional inter-dataset scatter (model discrepancy).}
Even after the clipping step, the residual scatter of the data about a smooth trend can exceed
that expected from the quoted point-to-point uncertainties alone. I therefore introduce a single energy-independent systematic uncertainty
term, \(\sigma_{\mathrm{sys}}\), in log space and add it in quadrature:
\begin{equation}
\sigma_{y,i}^{\mathrm{eff}}=\sqrt{\sigma_{y,i}^2+\sigma_{\mathrm{sys}}^2}.
\label{eq:sigmasc_global}
\end{equation}
For all datasets except the inverse-kinematics data of
Ref.~\cite{NanWang}, the uncertainty assignment is based on a fixed
conservative fractional component,
\[
\delta_{\rm sys}=0.25,
\]
added in quadrature to the quoted point-to-point uncertainties, as
defined in Eq.~(\ref{eq:delta_eff}).
\paragraph{How the best-fit curve and its uncertainty are obtained .}
I model the smooth trend of the data in log space by writing
$y(E)=\log_{10}S(E)$ as a quadratic function of energy about a pivot point $E_0$,
\[
y(E)=a_0+a_1(E-E_0)+a_2(E-E_0)^2,
\]
so the three numbers $(a_0,a_1,a_2)$ determine the entire curve. 
For each measured point $(E_i,y_i)$ we compare the model prediction $y(E_i)$ with the
measured value $y_i$, and I weight this difference by the corresponding uncertainty
$\sigma_{y,i}^{\rm eff}$. Points with smaller uncertainties have more influence on the fit,
while points with larger uncertainties have less influence. The ``best-fit'' parameters are
those that make the weighted sum of squared differences between the model and the data as
small as possible (this is the standard weighted least-squares criterion).

Assuming the deviations of the data from the smooth curve are approximately Gaussian in
$\log_{10}S$, and adopting a non-informative (flat) prior for the parameters, the resulting
probability distribution for the fitted coefficients $(a_0,a_1,a_2)$ is a multivariate normal
distribution centered on the best-fit values. In this case the most probable parameter set
(the maximum-a-posteriori estimate) coincides with the best-fit values. I then propagate
this parameter distribution to obtain the MAP curve and the credible bands for $S(E)$.
\paragraph{Weighted least squares and Gaussian posterior.}
Let $y=(y_1,\dots,y_N)^T$ and define the design matrix row
\begin{equation}
X_i = \bigl(1,\ E_i-E_0,\ (E_i-E_0)^2\bigr),
\end{equation}
so that $y(E_i)=X_i\cdot a$. With weights
$W=\mathrm{diag}\bigl(1/\sigma_{y,i}^{\mathrm{eff}\,2}\bigr)$,
the weighted least-squares solution is
\begin{equation}
a_{\mathrm{WLS}}=(X^T W X)^{-1}X^T W y.
\label{eq:awls_global}
\end{equation}
Assuming a Gaussian likelihood and a flat prior in $a$, the posterior for $a$ is multivariate
normal,
\begin{equation}
p(a|D)=\mathcal{N}(a_{\mathrm{mean}},\,C_a),\qquad
a_{\mathrm{mean}}=(X^T W X)^{-1}X^T W y,\qquad
C_a=(X^T W X)^{-1}.
\label{eq:posterior_a_global}
\end{equation}
For a multivariate normal posterior, the MAP estimator coincides with the posterior mean:
$a_{\mathrm{MAP}}=a_{\mathrm{mean}}$.

\subsection{Construction of the MAP curve and the 68\% credible band.}
I draw samples $a^{(k)}\sim p(a|D)$ and map each sample to
\begin{equation}
S^{(k)}(E)=10^{\,a^{(k)}_0+a^{(k)}_1(E-E_0)+a^{(k)}_2(E-E_0)^2}.
\label{eq:SEsamples_global}
\end{equation}
At each energy on a dense grid, the 68\% pointwise credible band is given by the
16th and 84th percentiles of the sample ensemble $\{S^{(k)}(E)\}$, while the central curve
is the MAP prediction
$S_{\mathrm{MAP}}(E)=10^{\,a_{\mathrm{MAP}}\cdot(1,E-E_0,(E-E_0)^2)}$.
(For reference, the median is the 50th percentile; we plot MAP to be consistent with the
Gaussian posterior for $a$.)

\end{document}